\documentclass[twocolumn,3p,preprint]{elsarticle}
\journal{Physics Letters B}
\usepackage[numbers]{natbib}
\RequirePackage{fancyhdr}
\RequirePackage{graphicx}
\usepackage{placeins}
\RequirePackage{adjustbox}
\usepackage{rotating}
\usepackage{multirow}
\usepackage{geometry}
\usepackage{ragged2e}
\usepackage{color}
\usepackage{array}
\usepackage{float}
\usepackage{units} 
\RequirePackage{amssymb}
\RequirePackage{epstopdf}
\usepackage[pdftex,colorlinks,linkcolor = blue,urlcolor  = blue,citecolor = blue]{hyperref}
\RequirePackage{amsmath, amsfonts}
\biboptions{sort&compress}
\bibpunct{[}{]}{,}{n}{,}{,}

\begin{document}
\begin{frontmatter}

\title{Strong enhancement of level densities in the crossover from spherical to deformed neodymium isotopes}

\author[UiO]{M.~Guttormsen}\ead{m.s.guttormsen@fys.uio.no}
\author[Yale]{Y.~Alhassid}\corref{cor1}\ead{yoram.alhassid@yale.edu}
\author[Yale]{W.~Ryssens}
\author[EOU]{K.O.~Ay}
\author[EOU]{M.~Ozgur}
\author[EOU,STU]{E.~Algin}
\author[UiO]{A.C.~Larsen}
\author[UiO]{F.L.~Bello~Garrote}
\author[UiO]{L.~Crespo Campo}
\author[UiO]{T.~Dahl-Jacobsen}
\author[UiO]{A.~G{\"o}rgen}
\author[UiO]{T.W.~Hagen}
\author[UiO]{V.W.~Ingeberg}
\author[UiO,UJ]{B.V.~Kheswa}
\author[UiO]{M.~Klintefjord}
\author[UiO]{J.~E.~Midtb{\o}}
\author[UiO]{V.~Modamio}
\author[UiO]{T.~Renstr{\o}m}
\author[UiO]{E.~Sahin}
\author[UiO]{S.~Siem}
\author[UiO]{G.~M.~Tveten}
\author[UiO]{F.~Zeiser}

\address[UiO]{Department of Physics, University of Oslo, N-0316 Oslo, Norway}
\address[Yale]{Center for Theoretical Physics, Sloane Physics Laboratory,
Yale University, New Haven, Connecticut 06520, USA}
\address[EOU]{Department of Physics, Eskisehir Osmangazi University, Faculty of Science and Letters, TR-26040 Eskisehir, Turkey}
\address[STU]{Department of Electrical and Electronics Engineering, AAT Science and Technology University, 01250 Adana, Turkey}
\address[UJ]{Department of Physics, University of Johannesburg, P.O.~Box 524, Auckland Park 2006, South Africa}

\cortext[cor1]{Corresponding author}
 
\begin{abstract}

Understanding the evolution of level densities in the crossover from spherical to well-deformed nuclei has been a long-standing problem in nuclear physics.  
We measure nuclear level densities for a chain of neodymium isotopes $^{142,144-151}$Nd which exhibit such a crossover. 
These results represent to date the most complete data set of nuclear level densities for an isotopic chain between neutron shell-closure and towards mid-shell.
We observe a strong increase of the level densities along the chain with an overall increase by a factor of $\approx 170$ at an excitation energy of 7.5 MeV and saturation around mass 150.  
Level densities calculated by the shell model Monte Carlo (SMMC) are in excellent agreement with these experimental results. 
Based on our experimental and theoretical findings, we offer an explanation of the observed mass dependence of the level densities in terms of the intrinsic single-particle level density and the collective enhancement.

\end{abstract}

\begin{keyword}
Nuclear level density, Oslo method, shell model Monte Carlo, mean-field theory, collective enhancement
\end{keyword}

\end{frontmatter}

\section{Introduction}
\label{sec:intro}

Compound-nucleus reaction cross sections are indispensable in a variety of applications such as understanding stellar nucleosynthesis~\cite{Arnould07}, designing next-generation nuclear reactors~\cite{Aliberti06}, and optimizing transmutation of nuclear waste~\cite{Aliberti04}. 
Such reactions are well understood by Hauser-Feshbach theory~\cite{Hauser52}, but this theory requires as input statistical nuclear properties, such as the nuclear level density (NLD).  
However, the microscopic calculation of NLDs in the presence of correlations is a challenging many-body problem.   
Furthermore, experimental data are usually limited to low excitation energies~\cite{ENSDF} and neutron resonance measurements at the neutron separation energy~\cite{RIPL-3}, making it difficult to benchmark theoretical models.  

Understanding the effects of deformation on the NLD is a long-standing open problem in nuclear physics;
see, \textit{e.g.}, Ref.~\cite{dossing2019} and references therein.
Compared to a spherical nucleus of similar mass, the NLD
of a deformed nucleus is determined by two competing effects: 
(i) in mean-field theory, the onset of deformation breaks
the magnetic degeneracy of the spherical single-particle levels, leading to an effective decrease of the average single-particle level density at the Fermi energy~\cite{Strutinsky77}, and thus lowers the NLD of intrinsic states; 
and (ii) rotational bands built on top of each of these intrinsic configurations lead to enhancement of the NLD~\cite{Bjornholm73,BM69}.  

Bj{\o}rnholm \textit{et al.}~\cite{Bjornholm73} predicted a collective enhancement factor $\approx 10$ for vibrations and $\approx 100$ for rotations, altogether a factor of $\approx 1000$.  Various phenomenological NLD models~\cite{Bethe36,Gilbert65,Dilg73,Egidy05,Koning08,Capote2009} were modified to take into account these enhancement factors. 
Yet modern combinatorial and mean-field methods must still have to be augmented by phenomenological collective enhancement factors~\cite{Hilaire06,Hilaire12,Uhrenholt13}. Shell effects and pairing correlations were included empirically in an energy-dependent level density parameter~\cite{ignatyuk1979}.

Recent shell-model Monte Carlo (SMMC) calculations~\cite{Ozen13,Alhassid2016} indicated a significantly smaller collective enhancement than suggested by Bj{\o}rnholm \textit{et al.}~\cite{Bjornholm73}. However, as stated by Junghans \textit{et al.}~\cite{junghans1998}, experimental information on the NLD is mandatory for quantifying this effect. 
A reliable theoretical model should reproduce the NLD well not only at the low-lying discrete levels and at the neutron resonance energy, but also for a broad excitation-energy region. The Oslo method used in this work provides the functional form of the NLD in the energy range between the low-lying discrete levels and the neutron separation energy, where often there is no other experimental data available.

We present a systematic study of the NLD for a chain of neodymium isotopes, starting from $^{142}$Nd at the $N=82$ shell closure and up to the well-deformed $^{151}$Nd, probing the effect of collectivity on the NLD in an unprecedented way.
We have also performed microscopic SMMC calculations for $^{142}$Nd up to $^{152}$Nd and find them to be overall in excellent agreement with experiment.
We show that the combined effect of a decrease in the single-particle level density with mass number and a collective enhancement results in an increase of the NLD with deformation that saturates around mass $150$.

\section{Experiment}
\label{sec:exp-results}
The light-ion reactions were performed at the Oslo
Cyclotron Laboratory. 
The targets were 
metallic foils of $^{142,144,146,148,150}$Nd with thicknesses of $\approx 2$~mg/cm$^2$ 
and enrichments of $\approx 97$\%. The targets were bombarded with proton and 
deuteron beams of energies 16.0 MeV and 13.5 MeV, respectively. 
The SiRi 
particle-telescope system~\cite{siri} was applied to determine the outgoing 
particle type and energy. 
The 64 particle telescopes were located $\approx$~5 cm
 from the target in eight angles between $126^{\circ}$ and $140^{\circ}$ with 
respect to the beam direction. 
The particle energy resolution was $\approx$150 
keV (FWHM). The $\gamma$ rays following the reactions were measured with 
the NaI(Tl) scintillator array CACTUS~\cite{CACTUS} and the LaBr$_3$(Ce) scintillator array 
OSCAR. Additional details are provided in the Supplemental Material~\cite{supplemental} .

\begin{figure*}[t]
\includegraphics[clip,width=2.0\columnwidth]{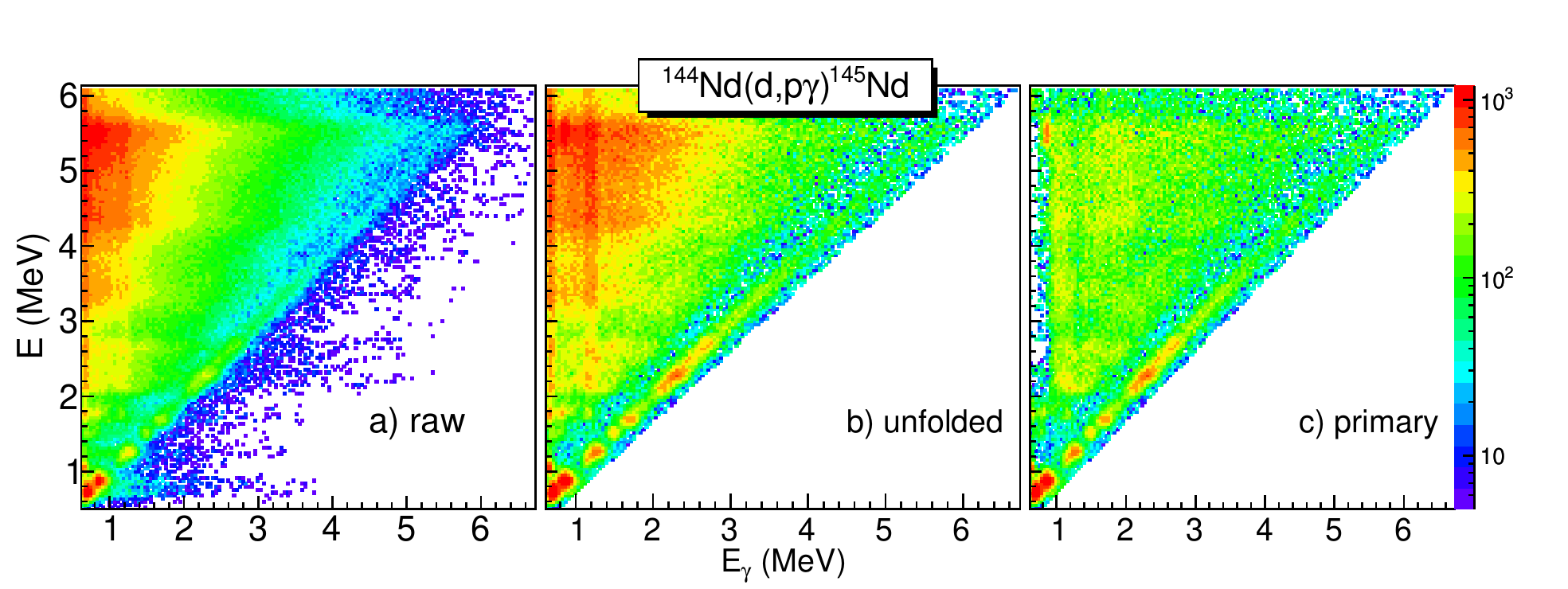}
\caption{Matrices with initial excitation energy $E$ versus $\gamma$-ray energy $E_{\gamma}$ from particle-$\gamma$ coincidences obtained by the $^{144}$Nd($d,p\gamma$)$^{145}$Nd reaction. The first steps of the Oslo method consist of establishing the a) raw, b) unfolded, and c) primary $\gamma$-ray matrices.}
\label{fig:matrices}
\end{figure*}

\section{The Oslo method}
\label{sec:oslo}

We extract the NLD for $^{142,144-151}$Nd applying the Oslo method for a set of particle-$\gamma$ ray coincidences. 
From the measured ejectile, we obtain information on the initial excitation energy $E$ of the residual nucleus. The $\gamma$ rays detected in coincidence with the ejectile 
reveal the decay properties from this specific excitation energy. 
Figure \ref{fig:matrices}a shows how the data are sorted into a matrix of 
initial excitation energies $E$ versus the $\gamma$-ray energy $E_\gamma$. This raw 
matrix is unfolded for each excitation energy bin (Fig.~\ref{fig:matrices}b) using the known detector response 
functions~\cite{Gut96,om2020}. 
Finally, the first generation (primary) $\gamma$-ray matrix $P$($E_\gamma,E$) is
obtained, see Fig.~\ref{fig:matrices}c. The first-generation procedure is based on an iterative 
subtraction technique~\cite{Gut87} which separates the distribution of the
first emitted $\gamma$ rays from all available $\gamma$ cascades.

The next step in the Oslo method is to factorize the primary $\gamma$-ray matrix by
\begin{equation}
P(E_{\gamma},E)\propto \mathcal{T}(E_\gamma)\rho(E-E_\gamma).
\label{eq:fac}
\end{equation}
Here, we have applied the Brink hypothesis~\cite{brink}: the $\gamma$-ray transmission coefficient $\mathcal{T}$ is approximately independent of excitation energy and spin/parity. 
The factorization is justified by Fermi's golden rule~\cite{dirac,fermi}, which states that the decay rate is proportional to the NLD at the final excitation energy after emitting the primary $\gamma$ ray. 
The fitting procedure performed using Eq.~(\ref{eq:fac}) enables the simultaneous extraction the NLD and the $\gamma$-ray transmission coefficient. However, it has been shown~\cite{Schiller00} that any transformation of the form 
\begin{equation}
\rho(E-E_{\gamma}) \to A\exp[\alpha(E-E_{\gamma})]\rho(E-E_{\gamma}),
\label{eq:trans}
\end{equation}
gives the same fit to $P$($E_\gamma,E$). To determine the parameters $A$ and $\alpha$ in (\ref{eq:trans}), we use  other experimental data. 
At low excitations, we normalize the NLD to known discrete levels~\cite{NNDC}. At high excitations, we used the measured average neutron 
$s$-wave resonance spacing $D_0$~\cite{MugAtlas} at the neutron separation energy $S_n$. 

\begin{table*}[bth]
\centering
\caption{Parameters for extracting NLD and systematic uncertainties in neodymium isotopes. Also listed are the quadrupole deformation $\beta_2$ and temperature $T_{\rm CT}$ of Eq.~(\ref{eq:ct}).}
\begin{tabular}{cllllll}
\hline
\hline
A   &$\beta_2$   &$T_{\rm CT}$&$S_n$& $\sigma(S_n)$ &$D_0$           & $\rho(S_n)$ \\
    &            & (MeV)  &(MeV)& RMI              & (eV)           &(10$^6$MeV$^{-1})$ \\
\hline
142 &0.092(2)    &0.65(5)&9.828 &  6.6(7)            &19(4)$^c$     &    1.23(35)$^b$\\
143 &0.109(5)$^a$&0.61(3)&6.124 &  6.1(6)            &1035(135)      &    0.07(2)\\
144 &0.125(2)    &0.63(3)&7.817 &  6.3(6)            &37.6(21)       &    0.32(5)\\
145 &0.138(5)$^a$&0.59(3)&5.755 &  5.9(6)            &450(50)        &    0.16(4)\\
146 &0.151(2)    &0.62(3)&7.565 &  6.2(6)            &17.8(7)        &    0.67(11)\\
147 &0.176(5)$^a$&0.57(3)&5.292 &  5.8(6)            &346(50)        &    0.20(5)\\
148 &0.200(2)    &0.59(3)&7.333 &  6.1(6)            &5.9(11)        &    2.4(6) \\
149 &0.242(5)$^a$&0.54(3)&5.039 &  5.8(6)            &165(14)        &    0.42(9)\\
150 &0.283(2)    &0.61(4)&7.376 &  6.2(6)            &3.0(10)$^c$    &    4.8(18)$^b$\\
151 &0.314(10)$^a$&0.54(3)&5.335 & 6.0(6)            &169(11)        &    0.43(9)\\
152 &0.345(9)    &       &7.278 &  6.3(6)            &         &     \\
\hline
\hline
\end{tabular}
\label{tab:nld_parameters}

$^a$Interpolated between even-$A$ neighbors. $^b$Scaled from systematics~\cite{supplemental}. $^c$Adjusted to reproduce $\rho(S_n)$.
\end{table*}

 To convert the measured $D_0$ to total level density, we use the spin cutoff model~\cite{Bethe37,Ericson60}.  The values $\sigma(S_n)$ of the spin cutoff parameter at the neutron separation energy $S_n$ are estimated based on a rigid-body moment of inertia (RMI)~\cite{supplemental} and are tabulated in Table~\ref{tab:nld_parameters}.  Table~\ref{tab:nld_parameters} also includes the quadrupole deformation $\beta_2$  from Ref.~\cite{pritychenko2016} and the temperature $T_{\rm CT}$ extracted by fitting the constant-temperature formula
\begin{equation}
\rho_{\text{CT}}(E)=(1/T_{\rm CT}) \exp{[(E-E_0)/T_{\text{CT}}]}
\label{eq:ct}
\end{equation}
to the high-energy data points, where $E_0$ is a shift parameter to match $\rho(S_n)$~\cite{supplemental}.

\begin{figure*}[t]
\includegraphics[clip,width=2.1\columnwidth]{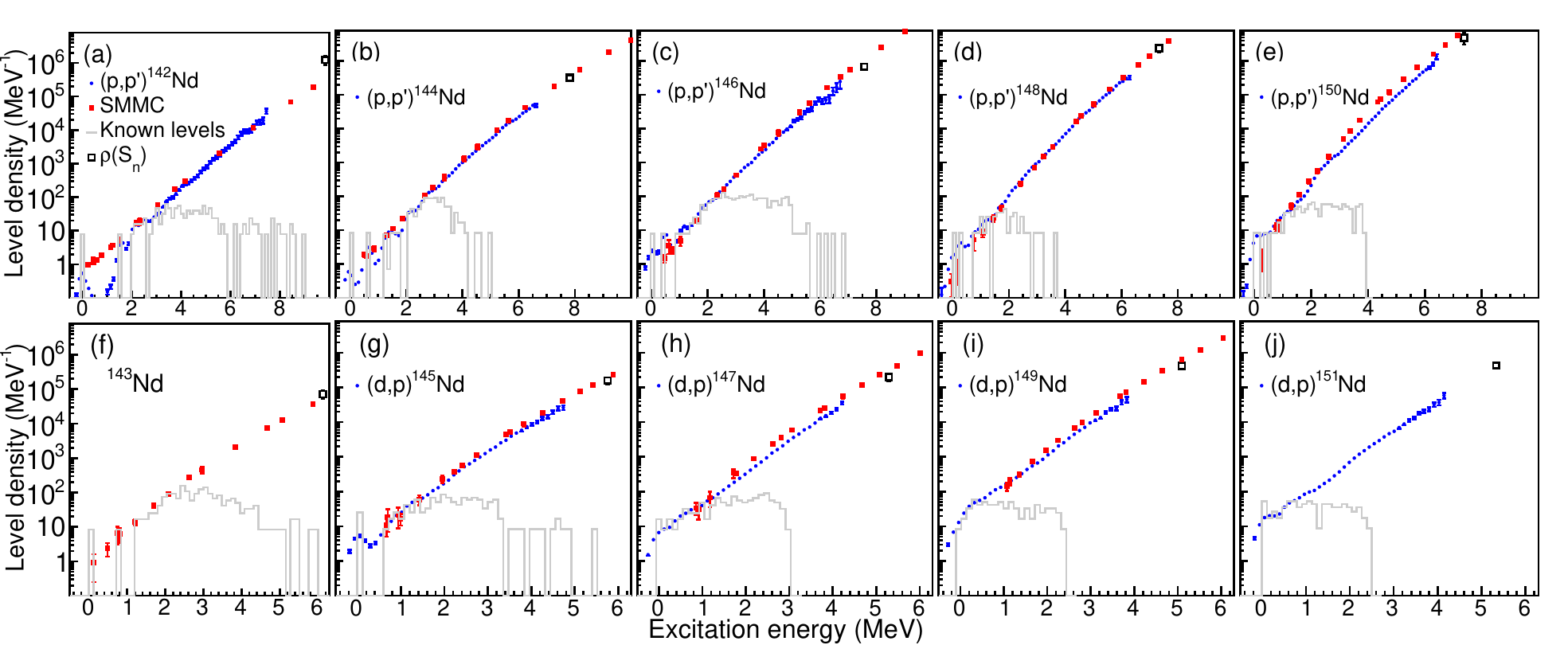}
\caption{Experimentally extracted NLDs (solid blue circles) of the $^{142,144-151}$Nd 
         isotopes. The gray histograms show the NLD of known discrete levels.
         The total NLDs evaluated from neutron capture resonance spacings  $D_0$
         are displayed as open black squares.  SMMC level densities for the $^{142-150}$Nd isotopes are shown
         by solid red squares.}
\label{fig:rhotot10}
\end{figure*}

\section{SMMC calculations}
\label{sec:SMMC}

The SMMC method~\cite{Lang93,Alhassid94} enables the exact calculation (up to statistical errors) of NLDs
in the framework of the configuration-interaction (CI) shell model.
This method allows us to use many-particle model spaces that are many orders of magnitude larger than those that can be treated by conventional shell model methods~\cite{Alhassid17}. 
In contrast to combinatorial and mean-field approaches, the SMMC approach does not require any empirical enhancement factors, and is therefore a suitable approach for studying the deformation dependence of the NLD. 

We carried out SMMC calculations in the proton-neutron formalism~\cite{Alhassid08} for the chain of neodymium isotopes  $^{142-152}$Nd.  
The CI shell model space includes the complete $50-82$ shell plus the $1f_{7/2}$ orbital for protons, and the complete $82-126$ shell plus the $0h_{11/2}$ and $1g_{9/2}$ orbitals for neutrons.  The effective interaction parameters are given in Ref.~\cite{Ozen13}.  
For the odd-mass isotopes, there is a  sign problem associated with the projection on an odd number of neutrons at low temperatures and the ground-state energies were taken from Ref.~\cite{Ozen15}. The latter estimated ground-state energies for all the odd neodymium isotopes in the chain with the exception of $^{151}$Nd.

In contrast to state densities that count the $2J+1$ degeneracy of each level with spin $J$, the measured level densities count each such level only once. 
In SMMC, the level densities are obtained by projection on $M=0$ ($M=1/2$) for even-mass (odd-mass) nuclei~\cite{Alhassid07,Alhassid15}.  
SMMC state densities for the neodymium isotopes were presented in Ref.~\cite{Alhassid14}. 
 We provide more details for the SMMC calculations in the Supplemental Material~\cite{supplemental}.

\section{Results}
\label{sec:results}
In Fig.~\ref{fig:rhotot10}  we compare the experimentally extracted NLDs of 
$^{142,144-151}$Nd with the SMMC results.
Above an excitation energy of $\sim 2-3$ MeV, the experimental NLDs are almost linear in a  logarithmic scale 
and are well-described by the constant-temperature formula (\ref{eq:ct}). 
It was conjectured that this behavior emerges once the first pair of nucleons is
broken~\cite{CT2015,Moretto2014,Moretto2015}, i.e., for an excitation energy 
$E > 2\Delta$, where $\Delta$ is the pairing gap. 
In contrast to recent findings in $^{167,168,169}$Tm~\cite{Pandit18}, we do not 
observe any experimental or theoretical signatures of irregular bumps in the NLD
curves.

\begin{figure}[bt]
\includegraphics[clip,width=\columnwidth]{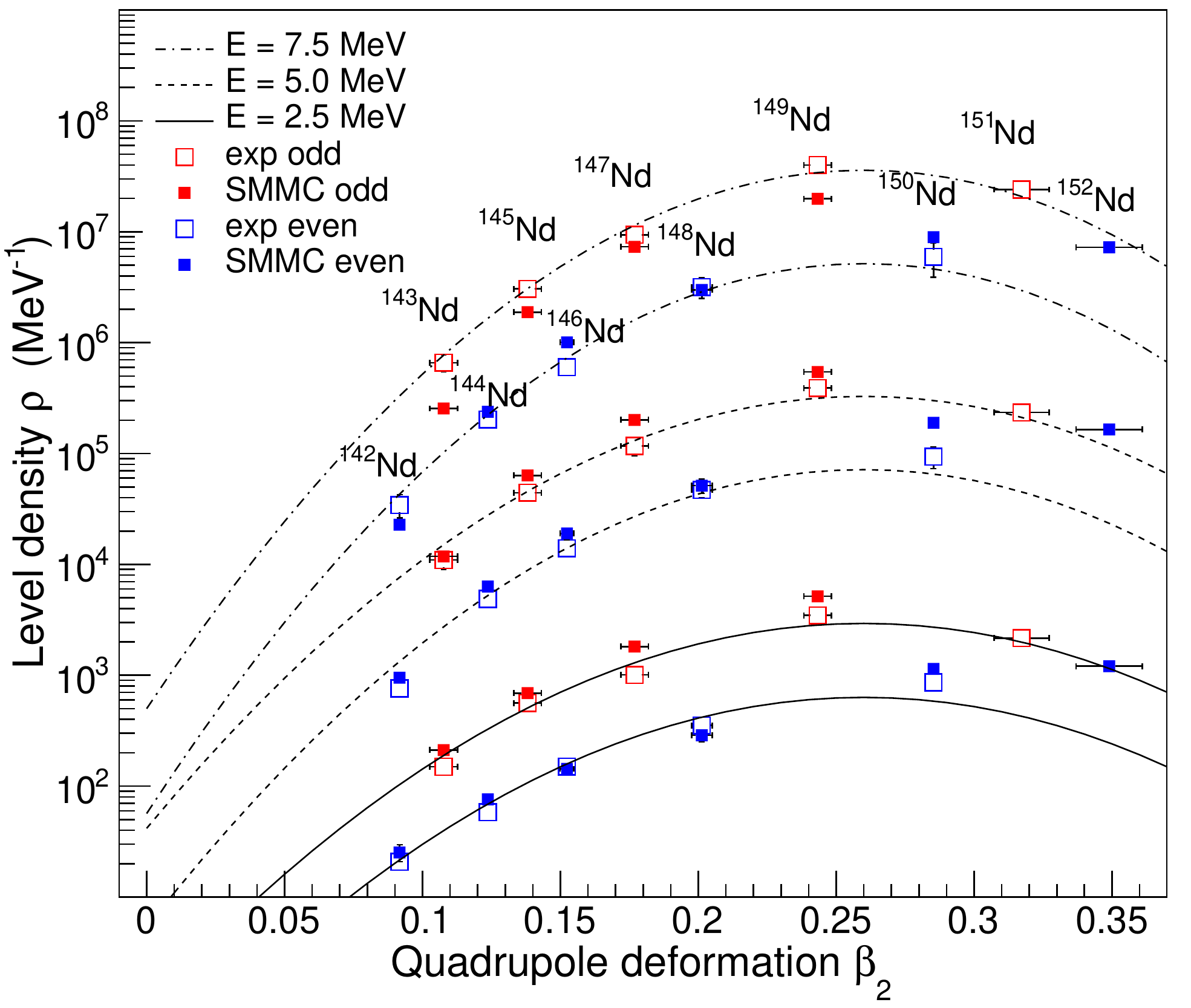}
\caption{Experimental (open squares) and SMMC (solid squares) level densities for $^{142-151}$Nd at excitation energies $E=2.5, 5.0$  and $7.5$~MeV. The experimental data points at $E=7.5$~MeV are extrapolated using the constant-temperature formula (\ref{eq:ct}) with values of $T_{\rm CT}$ given in Table~\ref{tab:nld_parameters}. The curves are calculated from Eq.~(\ref{eq:beta2-fits}); see text.}
\label{fig:rho_beta}
\end{figure}

Figure~\ref{fig:rho_beta} shows the experimental and SMMC NLDs from Fig.~\ref{fig:rhotot10} at three 
excitation energies of $2.5, 5$ and $7.5$ MeV as a function of deformation $\beta_2$. The deformation of 
the even-mass isotopes is determined from the compilation of Pritychenko 
{\em et al.}~\cite{pritychenko2016},
using the measured 
$B(E2)$ values between the ground state and the first excited $2^+$ state. 
For the odd-mass isotopes, we assume a deformation that is the average of their
even-mass neighbors.  These values of $\beta_2$  are listed in Table~\ref{tab:nld_parameters}; see also the Supplemental Material~\cite{supplemental}.

At excitation energies of $2.5$~MeV and $7.5$ MeV, the NLD is determined, respectively, by a complete set of known low-lying discrete levels and by the average neutron resonance spacing $D_0$, 
while at the intermediate excitation energy of $5$~MeV, the NLD is determined by the Oslo method.  We find that the deformation dependence of the experimental NLDs at these three excitation energies follow closely the empirical form
\begin{equation}
   \rho(\beta_2)=C\exp[-\eta(\beta_2-\beta_2^\mathrm{max})],
\label{eq:beta2-fits} 
\end{equation} 
where $C$ and $\eta$ are fit parameters and $\beta_2^\mathrm{max} = 0.25$. The resulting fits of Eq.~(\ref{eq:beta2-fits}) to the experimental data are shown by the curves in Fig.~\ref{fig:rho_beta}. We obtain similar values of the parameter $\eta$ for the even- and odd-mass isotopes with $\eta=118, 136$ and $166$ at $E = 2.5, 5.0$ and 7.5 MeV, respectively.  There is a strong odd-even effect where the NLD of an odd-mass nucleus is higher than the NLDs of its even-mass neighbors, 
which can be attributed to the blocking effect of the odd neutron~\cite{Ericson59}.

We observe overall excellent agreement between the experimental and SMMC NLDs. In particular, the trend with mass number is  well reproduced for both the even- and odd-mass isotopes.  For $^{152}$Nd there are no experimental results to compare with, but the SMMC results deviate from the curves fitted to the experimental systematics.  
In general, we find larger deviations for the odd-mass isotopes, in particular at the highest energy of $E=7.5$ MeV. 
We note, however, that the experimental results at this energy are estimated by extrapolating the Oslo data beyond the neutron separation energy using the constant-temperature formula \eqref{eq:ct}, and thus the comparison is not as conclusive as for the even isotopes. 

\begin{figure}[bt]
\includegraphics[clip,width=\columnwidth]{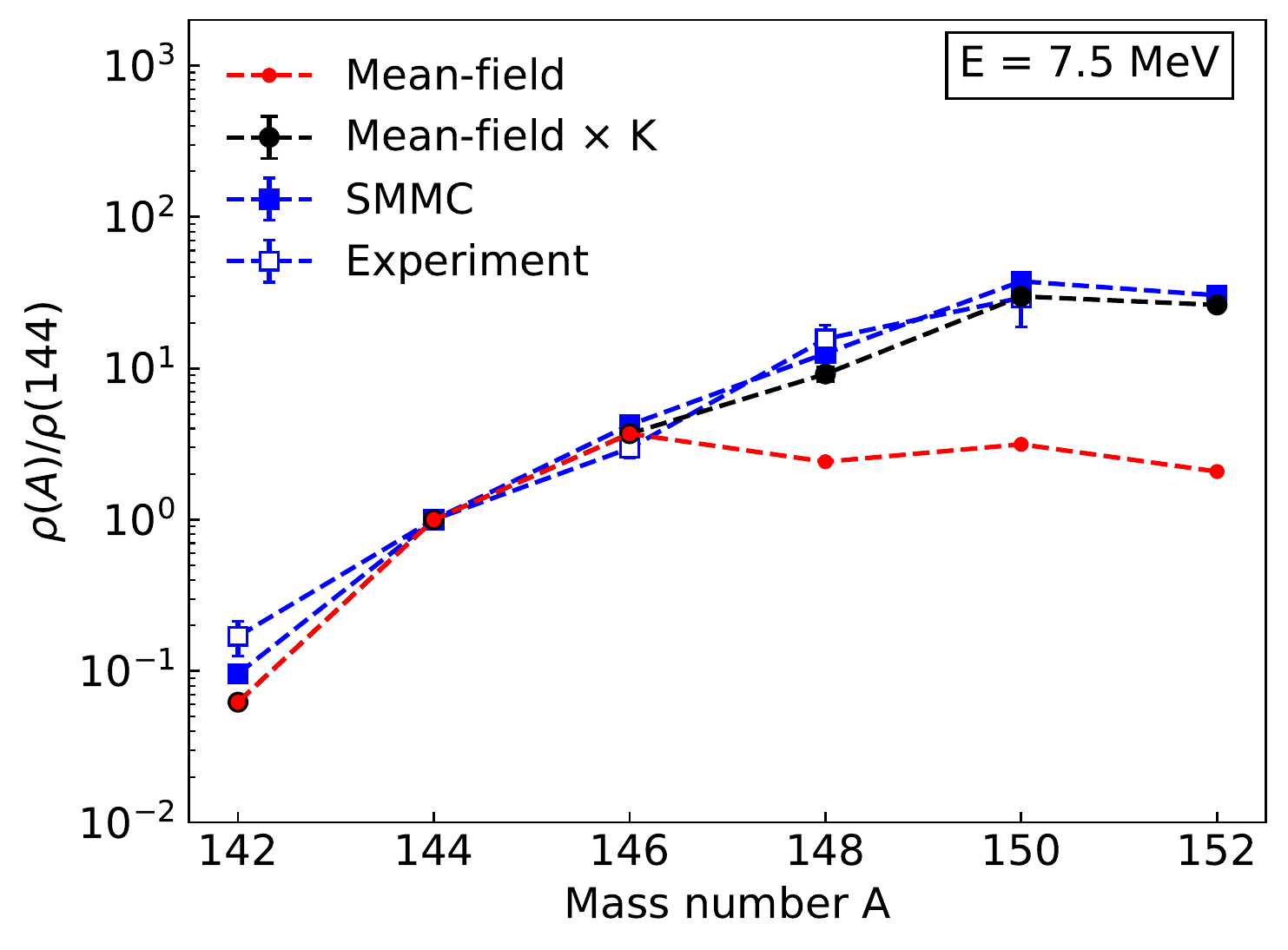}
\caption{The total enhancement $\rho(A)/\rho(A=144)$ of the NLDs of the even-mass neodymium isotopes relative to $^{144}$Nd at $E=7.5$~MeV. This enhancement, as determined from experiment (open blue squares), is compared with SMMC (solid blue squares), mean field (red circles) and the mean-field corrected by the collective enhancement factor $K$ (black circles). $K \approx 3.8, 9.5, 12.6$ in $^{148,150,152}$Nd, respectively.
}
\label{fig:ratio}
\end{figure}


\section{Explanation of the mass dependence}
 Figure~\ref{fig:ratio} shows the total enhancement of the NLD at $E= 7.5$ MeV of the even-mass neodymium isotopes relative to the NLD of $^{144}$Nd, the lightest isotope in the chain for which an experimental value of $D_0$ exists. Overall, we observe excellent agreement between experiment and the SMMC results.  We find a large enhancement in the experimental NLD by a factor of $\approx 170$ for $^{150}$Nd (relative to $^{142}$Nd). 
 
 The intrinsic mean-field state density  $\rho_{\rm mf}$, which is calculated using the same effective interaction as in SMMC,  
 is affected mostly by the average single-particle level density $g_n(\epsilon_F)$ of the neutrons at the Fermi energy $\epsilon_F$. 
 In deformed nuclei, rotational bands built on top of intrinsic band heads lead to an enhancement of the total state density, described by a collective enhancement factor $K=\rho_{\rm state}/\rho_{\rm mf}$~\cite{Ozen13}. 
 This collective enhancement of $\rho_{\rm state}$ is also reflected in the total level density, since these two densities are related by a spin cutoff parameter that is only weakly dependent on mass. 
 Thus, the mass dependence of the NLD is determined by two factors: $\rho_{\rm mf}$ and $K$.  With 82 neutrons, $^{142}$Nd is semi-magic and its neutron Fermi energy is in the middle of the shell gap and is thus characterized by a relatively low $g_n(\epsilon_F)$.  
 As we start filling the $82-126$ major shell in $^{144}$Nd,  the neutron Fermi energy rises, so that it is close to the $2f_{7/2}$ orbital, leading to a sharp increase in $g_n(\epsilon_F)$ and thus in $\rho_{\rm mf}$.  
 The $g_n(\epsilon_F)$ of the spherical mean-field solution continues to increase in $^{146}$Nd, though at a more moderate rate. 
 The increase in $\rho_{\rm mf}$ is shown by the red circles in Fig.~\ref{fig:ratio}.  
 The mean-field solution for our shell model interaction becomes deformed starting in $^{148}$Nd. Deformation lifts the spherical degeneracy of the single-particle levels and decreases $g_n(\epsilon_F)$ in the deformed isotopes, leading to a decrease in $\rho_{\rm mf}$ compared to $^{146}$Nd. 
 This decrease is compensated by the rise of $K$ with deformation. 
 As a result, the total NLD enhancement increases with mass but saturates around $^{150}$Nd, for which the SMMC NLD is very similar to that in $^{152}$Nd.  This is shown by the black circles in Fig.~\ref{fig:ratio}, describing the product of $K$ and the increase of $\rho_{\rm mf}$  (relative to $^{144}$Nd), and follow closely the observed total  enhancement  of the NLD with mass number  $A$.

\section{Conclusions}
\label{sec:conclusion}

 We extracted experimental NLDs for a long chain of neodymium isotopes using the Oslo method.
We observed a large total enhancement of the NLD in the crossover from spherical to deformed isotopes, which saturates around  $A=150$. 
The availability of experimental data for such a long isotopic chain makes these isotopes an excellent benchmark for testing the quality of current and future NLD models.
 We calculated SMMC NLDs of these isotopes in the framework of the CI shell model and found them to be overall in excellent agreement with the experimental NLDs.  
We explained the mass dependence of the NLDs by the combined effects of the intrinsic single-particle level density and of the collective enhancement.

\section*{Acknowledgements}
We thank J.~C.~M\"uller, P.~Sobas and J.~Wikne for
providing excellent experimental conditions. 
This work is supported by The Scientific and Technological Research Council of Turkey 
(TUBITAK) with project number 115F196. 
A.~C.~L. acknowledges funding of this
research by the European Research Council through ERC-STG-2014, grant 
agreement no. 637686. A.~C.~L. acknowledges support from the “ChETEC” COST 
Action (CA16117), supported by COST (European Cooperation in Science and 
Technology). 
This work benefited from support by the National Science Foundation
under Grant No. PHY-1430152 (JINA Center for the Evolution of the Elements).
This work was supported in part by the National Science Foundation under Grant No. OISE-1927130 (IReNA).
The work of Y.A. and W.R. was supported in part by the U.S.~DOE grant No.~DE-SC0019521. The SMMC calculations used resources of the National Energy Research Scientific Computing Center, which is supported by the Office of Science of the U.S. Department of Energy under Contract No.~DE-AC02-05CH11231.  We also thank the Yale Center for Research Computing for guidance and use of the research computing infrastructure.

\end{document}